\documentclass[aps,prl,twocolumn,superscriptaddress]{revtex4-1}
\usepackage{graphicx} 
\usepackage{amsbsy} 
\usepackage{amssymb,amsmath}
\usepackage{braket}
\usepackage{color}
\usepackage{ulem}
\usepackage{mathrsfs}

\begin{document}
\title{Landau quantization in coupled Weyl points: a case study of semimetal NbP}
\author{Y. Jiang}
\affiliation{School of Physics, Georgia Institute of Technology, Atlanta, Georgia 30332, USA}
\affiliation{National High Magnetic Field Laboratory, Tallahassee, Florida 32310, USA}
\author{Z. L. Dun}
\affiliation{School of Physics, Georgia Institute of Technology, Atlanta, Georgia 30332, USA}
\affiliation{Department of Physics and Astronomy, University of Tennessee, Knoxville, Tennessee 37996, USA}
\author{S. Moon}
\affiliation{National High Magnetic Field Laboratory, Tallahassee, Florida 32310, USA}
\affiliation{Department of Physics, Florida State University, Tallahassee, Florida 32306, USA}
\author{H. D. Zhou}
\affiliation{Department of Physics and Astronomy, University of Tennessee, Knoxville, Tennessee 37996, USA}
\author{M. Koshino}
\affiliation{Department of Physics, Osaka University, Toyonaka 560-0043, Japan}
\author{D. Smirnov}
\affiliation{National High Magnetic Field Laboratory, Tallahassee, Florida 32310, USA}
\author{Z. Jiang}
\email{zhigang.jiang@physics.gatech.edu}
\affiliation{School of Physics, Georgia Institute of Technology, Atlanta, Georgia 30332, USA}
\date{\today}

\begin{abstract}
Weyl semimetal (WSM) is a newly discovered quantum phase of matter that exhibits topologically protected states characterized by two separated Weyl points with linear dispersion in all directions. Here, via combining theoretical analysis and magneto-infrared spectroscopy of an archetypal Weyl semimetal, niobium phosphide, we demonstrate that the coupling between Weyl points can significantly modify the electronic structure of a WSM and provide a new twist to the protected states. These findings suggest that the coupled Weyl points should be considered as the basis for analysis of realistic WSMs.
\end{abstract}

\maketitle

Recently, WSMs have attracted great attention in the search of three-dimensional zero-gap materials with non-trivial band topology \cite{Hasan00,Dai0,Vafek1,BHYan0,Armitage1}. In a WSM, the conduction band (CB) and valence band (VB) touch at discrete points in momentum space\textemdash Weyl points (WPs). In the vicinity of these points, the electronic band structure can be described by an effective Hamiltonian resembling the Weyl equation in high energy physics. Consequently, the electrons in WSM not only exhibit a linear dispersion with momentum but also carry chirality, giving rise to many exotic physics such as the Fermi arcs \cite{Hasan01,Ding0,YLChen1,Hasan0}, the chiral anomaly effect \cite{ong1,Hasan0,BHYan01}, and the mixed axial-gravitational anomaly \cite{BHYan02}.

Since the WPs always come in pairs \cite{nogo}, the coupling between the two WPs can significantly modify the band dispersion, resulting in richer and unique responses. Specifically, when the linear band from each WP overlaps, the band symmetry is reduced from the spherical symmetry of isolated WP to the axial symmetry of two coupled WPs (CWPs). Band hybridization and avoided level crossing (anticrossing) are expected to occur. Recent studies have suggested that such a band modification is responsible for the breakdown of chiral anomaly and the opening of a notable band gap in high magnetic fields \cite{SJia1,Ramshaw01,Park1,Lee01,Rodionov1}. 

In this Letter, we combine theoretical calculations with infrared magneto-spectroscopy experiment to unveil the evidence of CWPs in niobium phosphide (NbP), belonging to the archetypal WSM family of nonmagnetic transition-metal monopnictides (TX: T=Ta, Nb; X=As, P) \cite{Hasan00,Dai0,Hasan01,Ding0}. The Landau level (LL) spectrum calculated in an experimentally relevant configuration, i.e., when the magnetic field ($B$) is applied perpendicular to the separation of the CWPs, reveals several unique spectral features that are directly associated with the coupling effects between the two WPs. We show that the magnetic field dependence of the observed inter-LL transitions deviates significantly from the expectation for isolated WPs, but can be consistently reproduced within the model of CWPs. Therefore, coupling between WPs should be considered when analyzing realistic WSMs.

\begin{figure}[b]
\includegraphics[width=8.5cm]{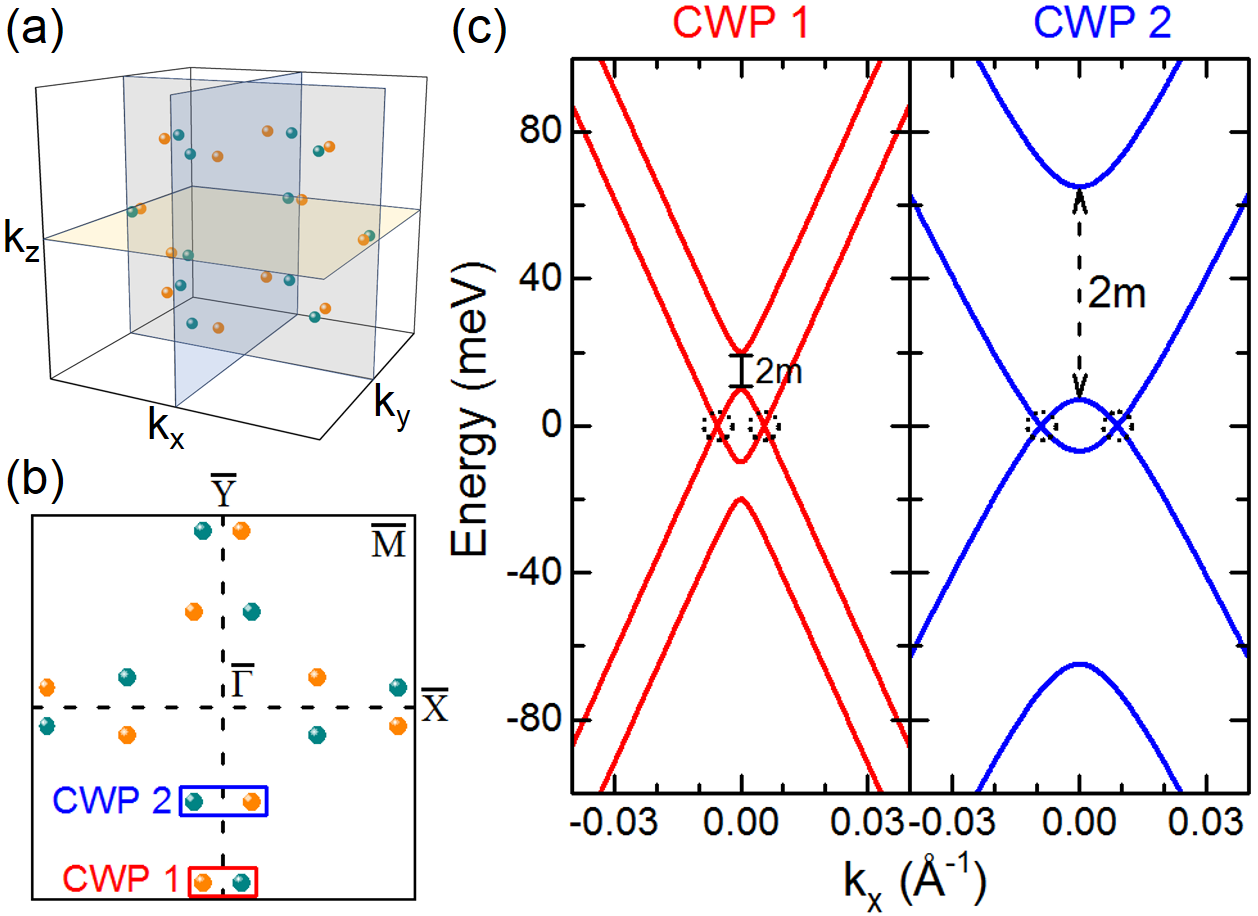}
\caption{(color online) (a) Schematic view of the 12 pairs of CWPs in NbP in $k$-space. (b) Schematic view of the CWPs projected onto the (001) surface Brillouin zone. In both (a) and (b), each CWPs consists of two WPs (ball symbols) with opposite chirality, color-coded in orange and dark cyan, respectively.  There are two types of CWPs, CWP 1 and 2, oriented either along the [100] or [010] direction, with the less coupled CWP 1 located just inside $\overline{X}$ (or $\overline{Y}$). (c) Low-energy dispersion of CWP 1 and 2, calculated using Eq. (\ref{zerobands}) with practical material parameters of NbP. For CWP 1, $v=3.9\times 10^5$ m/s, $m=5$ meV, and $b=15$ meV; for CWP 2, $v=3.6\times 10^5$ m/s, $m=29$ meV, and $b=36$ meV. The two WPs in each panel are enclosed by dot circles, and the $2m$ hybridization gap between the two CBs is indicated by the arrow.}
\label{Fig1}
\end{figure}

NbP as well as other nonmagnetic transition-metal monopnictide WSMs hosts 12 pairs of CWPs in the first Brillouin zone with their axial directions along [100] or [010] crystal axis (Figs. \ref{Fig1}(a,b)). The CWPs along [100] can be described by a $4 \times 4$ Hamiltonian \cite{Koshino1}
\begin{align}
\label{model}
H=v \tau_x\hbar (\boldsymbol{\sigma} \cdot \bold{k})+m\tau_z+b\sigma_x,
\end{align}
where $v$ is the band velocity, $\hbar\bold{k} = \hbar(k_x,k_y,k_z)$ is the momentum vector, and $\boldsymbol\sigma$ and $\boldsymbol\tau$ are the Pauli matrices for spin and pseudospin, respectively. The parameter $b$ quantifies the spin splitting in the material and $|b|>|m|$ (where $m$ is a mass parameter) is required to form a pair of CWPs. At zero magnetic field, Eq. (\ref{model}) leads to four energy bands
\begin{align}
\label{zerobands}
E_{s,\mu}=s\sqrt{m^2+b^2+(v\hbar \bold{k})^2+2\mu b\sqrt{(v\hbar k_x)^2+m^2} },
\end{align}
where $\mu = \pm 1$ and $s=\pm 1$. Examples of the low-energy dispersion are shown in Fig. \ref{Fig1}(c), using practical material parameters of NbP. One can identify the two WPs (enclosed by dot circles) in each panel of Fig. \ref{Fig1}(c) at $E=0$ and $\bold{k_W}=(\pm \frac{\sqrt{b^2-m^2}}{\hbar v},0,0)$. At $k=|\bold{k}|=0$, all energy bands reach a local extremum  with $E_{1,1}=b+m$ for the upper CB, $E_{-1,1}=-(b+m)$ for the lower VB, $E_{1,-1}=b-m$, and $E_{-1,-1}=-(b-m)$. The hybridization gap between the two CBs (VBs) is thus $2m$.

To calculate the LL spectrum, we assume the magnetic field is along the [001] direction, perpendicular to the separation of \textit{all} CWPs (i.e., $\bold{B}\perp\bold{k_W}$). This assumption, although different from the majority of existing literature (where $\bold{B}\parallel\bold{k_W}$ is assumed to preserves the axial symmetry), is relevant to experiment. This can be easily seen in Fig. \ref{Fig1}(b), where regardless of which crystal axis the magnetic field is applied along, $\bold{B}\perp\bold{k_W}$ is true for at least half of the CWPs.

\begin{figure}[h]
\includegraphics[width=8.6cm]{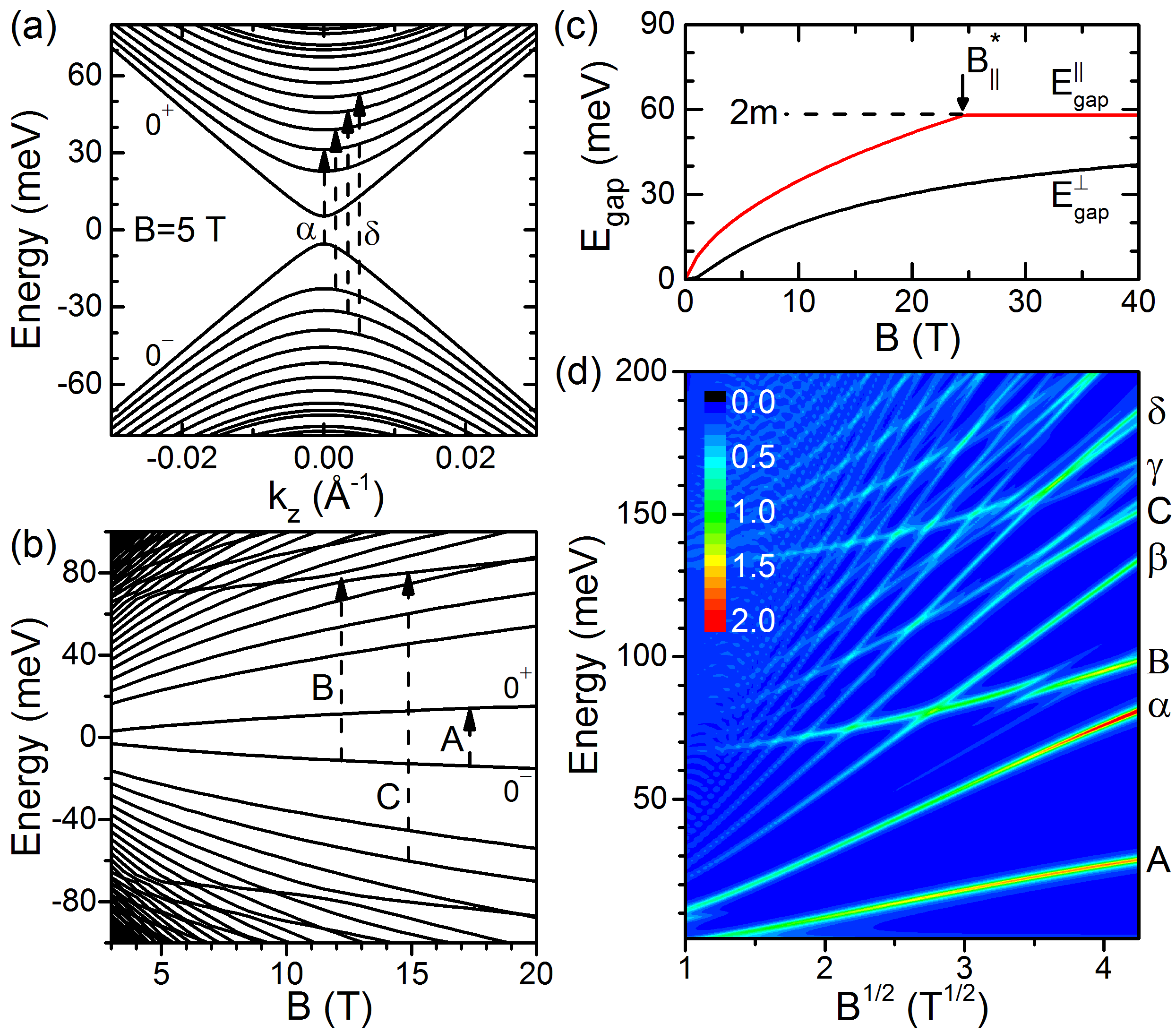}
\caption{(color online) (a) LL dispersion of CWPs along the magnetic field ($k_z$) direction at  $B=5$ T. (b) Magnetic field dispersion of the LLs at $k_z=0$. Dominant interband transitions are labeled by Greek letters ($\alpha, \beta, \gamma, \delta$), while transitions due to coupling between WPs are labeled by Roman letters (A, B, C). (c) Comparison of the band gap opening by the magnetic field applied perpendicular ($\bold{B}\perp\bold{k_W}$) or parallel ($\bold{B}\parallel\bold{k_W}$) to the WPs separation. In perpendicular geometry, the gap is always equal to the splitting of the $0^{\text{th}}$ LL (transition A). (d) False color plot of the real part of the optical conductivity at $k_z=0$, featuring a complex pattern of LL crossings/anticrossings. For demonstration purposes, the calculations are performed for CWP 2.}
\label{Fig2}
\end{figure}

Figures \ref{Fig2}(a,b) show the calculated LL dispersion along the magnetic field ($k_z$) direction and the magnetic field dependence of the LLs at $k_z=0$. The calculation was performed numerically due to the lack of analytical solution and details can be found in the Supplemental Material \cite{supp}. The calculated dispersion features a field-induced splitting of the $0^{\text{th}}$ LL, which cannot be explained within a single WP picture \cite{Carbotte1}. The magnetic field dependence of the gap energy, $E_{gap}^{\perp} = \Delta E_0(B_\perp)=E_{0^+}-E_{0^-}$, is plotted in Fig. \ref{Fig2}(c), where the gap increases gradually with $B_\perp$ and is expected to saturate at $\Delta E_0=2m$ in the high-field limit. This behavior is a signature of CWPs and can be deduced from the model above using perturbation theory. Specifically, in the high-field limit, $b \ll \Delta_B$ (where $\Delta_B=v\sqrt{2\hbar eB}$ is a $B$-dependent energy scale and $e$ is the elementary charge) and the Hamiltonian $H$ in Eq. (\ref{model}) is dominated by the first two terms with $H_1=b\sigma_x$ as a perturbation. Note that $H_0=v \tau_x\hbar (\boldsymbol{\sigma} \cdot \bold{k})+m\tau_z$ has a form of gapped graphene \cite{Zli1,Falko1} and its LL spectrum reads $E_n=\pm\sqrt{\Delta_B^2 n+m^2}$, where integer $n$ is the LL index and $\pm$ stands for the CB and VB, respectively. The first-order perturbation of $H_1$ leads to $\delta E^{(1)}_n=0$ for all $n$, therefore the gap $\Delta E_0$ is dominated by $H_0$. In the limit as $B\rightarrow \infty$, $\Delta E_0=2m$.

Another essential spectroscopic feature of CWPs is the interaction between the two CBs (VBs). Since the CB's (VB's) separation is relatively small, there exists a great number of intersections of the LLs belonging to different CBs (VBs). The band interaction results in a significant reconstruction of LLs. For example, one can easily notice in Fig. \ref{Fig2}(b) a series of crossings/anticrossings whenever the energy of the $n=0$ LL of the upper CV (lower VB) coincides that of the lower CB (upper VB) LLs. This behavior significantly modifies the spectrum of optical excitations of WSMs, as shown below in more detail.

\begin{figure*}[t]
\includegraphics[width=16cm]{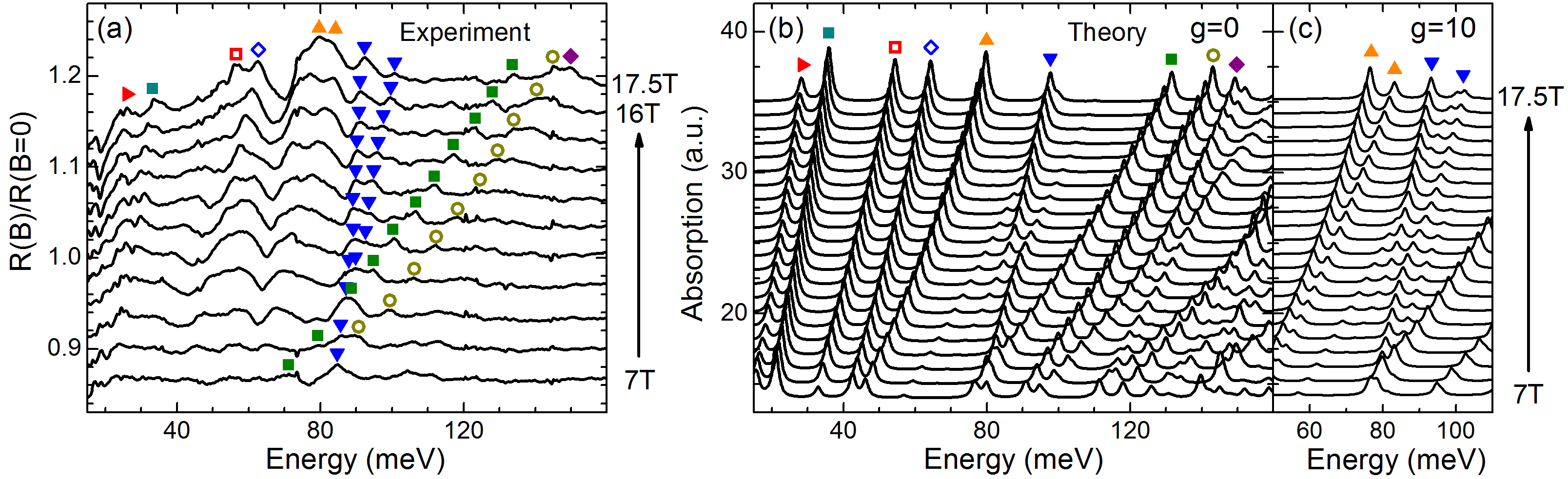}
\caption{(color online) (a) Normalized magneto-reflectance spectra, $R(B)/R(B=0)$, of NbP measured at selected magnetic fields. Dominant spectral peaks (or modes) can be grouped into several series marked by color-coded symbols along the 17.5 T curve. The color code is consistent with that used in Fig. \ref{Fig4}. The blue down-triangles indicate a peculiar mode that is weakly dependent on the magnetic field and crosses two ``fast'', strongly field-dependent, modes (olive squares and dark yellow circles) at $B\approx8$ T. (b) Calculated magneto-absorption spectra using the CWPs model and considering contributions from both types of CWPs (CWP 1 and 2). We attribute the modes marked by open symbols to CWP 1 and that of solid symbols to CWP 2. (c) Effect of additional symmetry breaking on stronger coupled CWP 2 accounted for by introducing an effective $g$-factor.}
\label{Fig3}
\end{figure*}

For an isolated WP (with Fermi energy $E_F=0$), the inter-LL transitions are graphene-like, following a selection rule of $\Delta n=\pm1$  \cite{Carbotte1,Carbotte2,ZJiang}. For CWPs, however, due to the band hybridization and LL reconstruction, the selection rules for optical transitions between LLs depend on the orientation of externally applied magnetic field with respect to the WP separation $2\bold{k_W}$. Figure \ref{Fig2}(d) shows the false color plot of the real part of the optical conductivity as a function of both energy and square root of the magnetic field for $\bold{B}\perp\bold{k_W}$. Two sets of transitions with relatively strong intensities can be identified. One set labeled by Greek letters is associated with dominant interband transitions, while the other set labeled by Roman letters consists of transitions specific to CWPs, either across $\Delta E_0$ (mode A) or involving anticrossing LLs (modes B and C). Because of electron-hole symmetry, each mode (except for mode A, $L_{0^-}$$\rightarrow$$L_{0^+}$) comprises two degenerate transitions, one electron-like ($\Delta n>0$) and the other hole-like ($\Delta n<0$). For demonstration purposes, only electron-like transitions and mode A are shown in Figs. \ref{Fig2}(a,b).

The interband transitions labeled by Greek letters are common features of Dirac and Weyl semimetals. However, due to the LL reconstruction, dominant transitions between the lower CB and the upper VB now follow a new selection rule of $\Delta n=\pm2$, as shown in Fig. \ref{Fig2}(a). We note that breaking of the usual selection rule $\Delta n=\pm1$ generally occurs when the energy band loses the axial symmetry. For instance, for a tilted WP, when the magnetic field is perpendicular to the tilt direction, more LL transitions are allowed including $\Delta n=\pm2$ \cite{Koshino_new,Yu_new,Udagawa_new,Goerbig_new}. The transitions labeled by Roman letters, on the other hand, are characteristics of CWPs. For example, the mode A does not exist for an isolated WP and is forbidden for a Dirac semimetal. Its high-field limit gives a direct measure of the mass parameter $m$. The mode B shows an interesting magnetic field dispersion. It first crosses the modes $\gamma$ and $\beta$, followed by an anticrossing type splitting, which is also indicated by the arrow in Fig. \ref{Fig2}(b). The zero-field limit of this mode gives an estimate of the band edge of the upper CB, $(b+m)$, or equivalently that of the lower VB, $-(b+m)$.

Next, we compare the CWPs model with our magneto-infrared reflectance measurements on NbP \cite{supp}. We choose NbP as the material platform for the following reasons. (1) NbP has the weakest spin-orbit coupling within its family \cite{YLChen1,BHYan03}, leading to relatively small values of $b$ and $m$ and easier access to the high-field limit. (2) In NbP, $E_F$ is very close to the WPs \cite{BHYan04,BHYan05}, i.e., near the intrinsic limit. (3) The trivial bulk bands in NbP are expected to quickly move away from the Weyl bands at a sufficiently low magnetic field \cite{BHYan04}, which minimizes the trivial contributions to the magneto-spectroscopy data.

\begin{figure}[t]
\includegraphics[width=8.5cm]{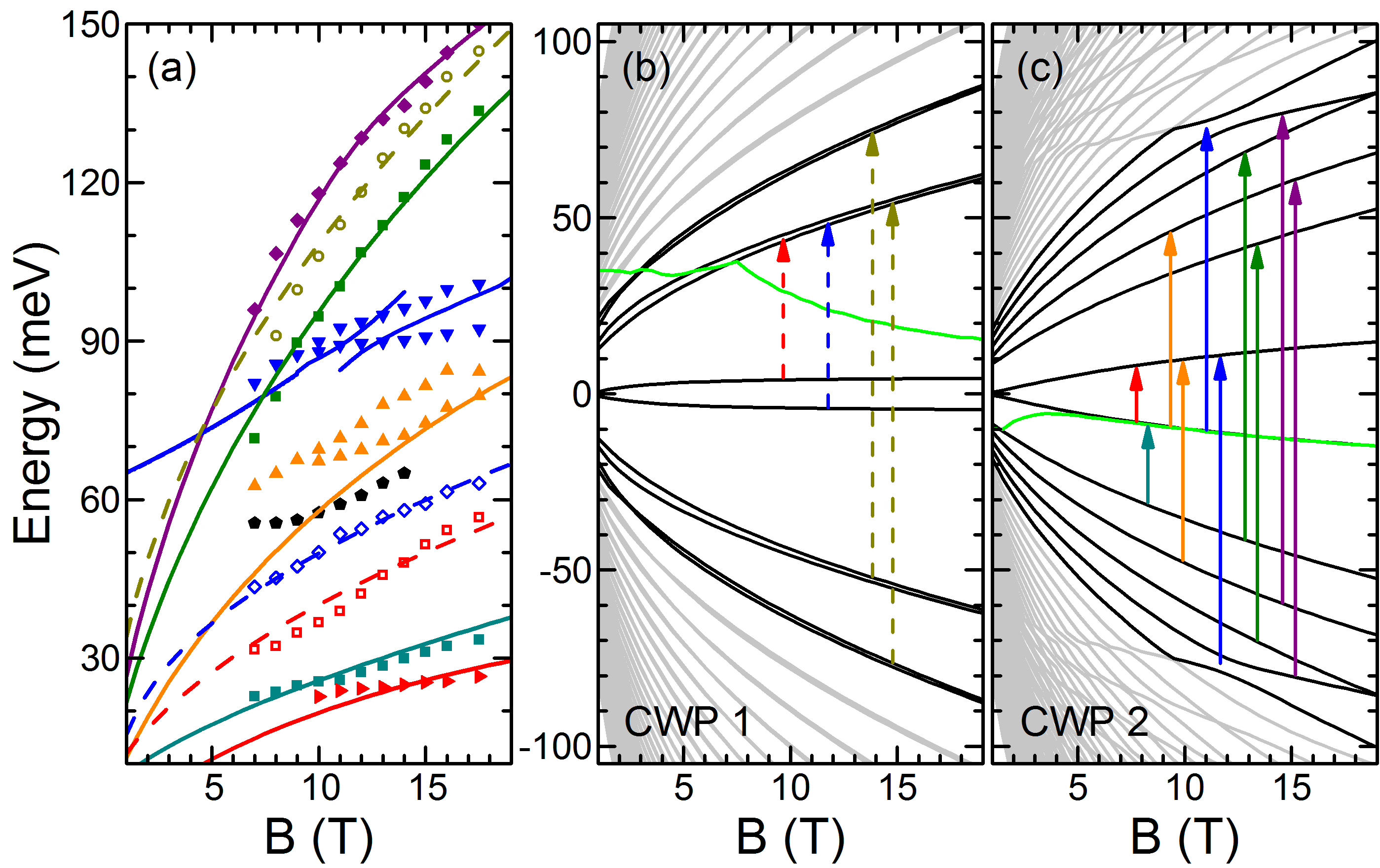}
\caption{(color online) (a) Magnetic field dependence of the major transitions observed in our experiment. The dash and solid lines are best fits to the data, when applying the CWPs model to CWP 1 and 2. The fits are color-coded and the corresponding transitions are shown in the LL fan diagram of CWP 1 (b) and 2 (c), where the low-lying LLs relevant to the observed transitions are plotted in black while the higher LLs are in grey. The green lines in (b) and (c) represent the expected Fermi levels as a function of magnetic field.}
\label{Fig4}
\end{figure}

Figure \ref{Fig3}(a) shows the normalized magneto-reflectance spectra, $R(B)/R(B=0)$, of NbP measured at selected magnetic fields ($\bold{B}\perp\bold{k_W}$). Dominant spectral peaks (or modes) can be grouped into several series marked by symbols along the 17.5 T curve. One can assign these modes to specific inter-LL transitions, with the transition energy approximated by the peak energy \cite{DDrew}. Figure \ref{Fig4}(a) summarizes the magnetic field dependence of the modes, where the low-energy ones generally exhibit a relatively weak $B$-dependence while the three higher-energy modes increase rapidly with magnetic field. Therefore, the mode guided by blue down-triangles inevitably crosses the fast increasing modes labeled by olive squares and dark yellow circles at $B\approx8$ T. Such a crossing behavior certainly cannot be explained by a single WP model, but consistent with the mode B or C in Fig. \ref{Fig2}(d) for CWPs.

It is instructive to estimate the relevant band parameters before a full theoretical calculation and quantitative fits to our data. First, we consider the contributions from both CWP 1 and 2 (Fig. \ref{Fig1}(c)), with the two WPs in CWP 1 closer to each other and having a smaller hybridization gap between the two CBs (VBs). This is in agreement with recent angle-resolved photoemission spectroscopy (ARPES) studies \cite{Ding1,Yando}. Second, we estimate the band velocity using the three fast increasing modes (with respect to magnetic field) observed in our experiment. These modes occur at relatively high energies and exhibit a $\sqrt{B}$-like dependence. The band velocity can be extracted using a method similar to that for gapped graphene \cite{Zli1,Falko1}. Third, we assign the lowest mode observed (labeled by right-triangles in Figs. \ref{Fig3}(a) and \ref{Fig4}(a)) as the mode A of CWP 2. As mentioned above, the fit to this mode determines the mass parameter $m$. Fourth, we assign the mode marked by down-triangles in Figs. \ref{Fig3}(a) and \ref{Fig4}(a) as the mode B of CWP 2. Extracting this mode to zero field leads to an estimate of $b+m$. Lastly, for CWP 1, since the two WPs are less coupled but nearly degenerated, it resembles the electronic structure of a Dirac semimetal (but with a field-induced gap/mass). The inter-LL transitions can then be analyzed accordingly \cite{Jiang1}.

Based on these estimations, we calculate the magneto-absorption spectra of NbP \cite{supp} and fit the low-lying modes to the data in Fig. \ref{Fig4}(a). The dash and solid lines are best fits to the data, when applying the CWPs model to CWP 1 and 2. The calculated LL fan diagrams and the corresponding inter-LL transitions are shown in Figs. \ref{Fig4}(b,c), respectively. Semi-quantitative agreement between the experiment and calculation is achieved using the band parameters $b=15$ meV, $m= 5$ meV, $v=3.9\times 10^5$ m/s, and $E_F=35$ meV (electron-doped) for CWP 1 and $b=36$ meV, $m=29$ meV, $v=3.6\times 10^5$ m/s, and $E_F=-10$ meV (hole-doped) for CWP 2. We note that the assigned $E_F$ values do not affect the LL spectrum of CWP 1 and 2, but Pauli-block certain transitions. The calculated magnetic field dependence of the $E_F$ is shown in Figs. \ref{Fig4}(b,c). The assumed $E_F$ values and the band parameters of CWP 1 and 2 are highly consistent with that reported in recent experimental and theoretical studies of NbP \cite{Ding1,BHYan3,Yando}. In Fig. \ref{Fig3}(b), we also plot the calculated total magneto-absorption spectra to provide a side-by-side comparison with the experimental data in Fig. \ref{Fig3}(a).

Finally, we discuss the splitting features in Fig. \ref{Fig4}(a) that remain unexplained within the present CWPs model. The presence of additional splitting is a signature of another symmetry breaking mechanism, which could be the lifting of the spin degeneracy and/or the breaking of the electron-hole (e-h) symmetry \cite{Jiang1}. Indeed, both ARPES measurements and first-principles calculations have shown an e-h asymmetric band structure in NbP \cite{Ding1}. The splitting caused by e-h asymmetry can be modeled in a similar way as the spin splitting \cite{Jiang1} via introducing an effective $g$-factor and adding a Zeeman-like term into the Hamiltonian of Eq. \ref{model}. The splitting is more pronounced for low-lying transitions that involve the $0^{\text{th}}$ LL.  As a result, the lowest interband transitions ($\alpha$ and B) split into two separate modes each (Fig. \ref{Fig3}(c)), giving rise to a more quantitative agreement with the experimental data as shown in Fig. S1 of the Supplemental Material \cite{supp}. The only major mode that cannot be explained with our CWPs model is the black pentagon mode in Fig. \ref{Fig4}(a). The optical weight of this mode exhibits an unusual magnetic field dependence, diminishing at $B\geq15$ T. We suspect that it originates from the trivial parallel bands in $k$-space away from the CWPs, as that suggested by recent transport and optical studies \cite{Wang_new,Neubauer_new}.

As a remark, we emphasize that the magnetic-field-induced gap is a hallmark of CWPs. The gap opening mechanism, however, is different between the case of $\bold{B}\perp\bold{k_W}$ and $\bold{B}\parallel\bold{k_W}$. When $\bold{B}\parallel\bold{k_W}$, the gap's origin is defined by the magnetic field strength as the $n=-1^+$ LL may sit above or below the $0^-$ LL with a crossing at $B_{\parallel}^*=\dfrac{2bm}{e\hbar v^2}$ (corresponding to $\Delta^2_B=4bm$ and see also in Fig. S3 \cite{supp}). For the case of $\bold{B}\perp\bold{k_W}$ considered here, the gap is always equal to the separation between the two $0^{\text{th}}$ LLs. The gap opening mechanism can be thought as Weyl annihilation \cite{SJia1} when the wavefunction broadening in $k$-space (which is $\sim1/l_B$, where $l_B=\sqrt{\hbar/eB}$ is the magnetic length) exceeds the separation of the two WPs, $2 k_W$. In the high-field limit, the gap reaches $2m$ as for a system of Weyl fermions with a finite relativistic mass of $m$. The band gap evolution with $B$ for both cases is given below and summarized in Fig. \ref{Fig2}(c).  
\begin{align}
\label{gap1}
E_{gap}^\perp&=E_{0^+}-E_{0^-}  \underset{B \rightarrow \infty}{ \xrightarrow{\hspace*{1cm}}} 2m,  \nonumber
\\E_{gap}^\parallel&=
\begin{cases} 
      E_{0^+}-E_{-1^+}=-(b-m)+\sqrt{(b-m)^2+\Delta_B^2},
      \\ \hspace{5.6cm} (B \leq B_{\parallel}^*)
      \\ E_{0^+}-E_{0^-}=2m, \hspace{2.8cm} (B > B_{\parallel}^*)
      \nonumber
   \end{cases}
\end{align}

In conclusion, we have demonstrated both theoretically and experimentally the essential role of the coupling effect between WPs in an established WSM, NbP. The combination of low Fermi energy and relatively small separation between WPs makes NbP an excellent platform for a case study of CWPs physics. Our band structure analysis predicted  several unique spectroscopic features originated from the CWPs that were largely confirmed in the magneto-spectroscopy experiment. These results emphasize the importance of coupling between WPs both for fundamental understanding of Weyl fermions in realistic condensed matter systems and for future device applications as well their limitations.

This work was primarily supported by the DOE (Grant No. DE-FG02-07ER46451). The NbP crystal growth at UT was supported by the NSF (Grant No. DMR-1350002). M.K. was supported by JSPS KAKENHI (Grant No. JP25107001, JP25107005, and JP17K05496). The infrared measurements were performed at the National High Magnetic Field Laboratory (NHMFL), which is supported by NSF Cooperative Agreement No. DMR-1157490 and the State of Florida. Z.J. acknowledges support from the NHMFL Visiting Scientist Program.

\end{document}